  \documentclass[12pt,preprint]{aastex}
 \usepackage{ifpdf}
 \usepackage{epsfig}
  \usepackage[latin1]{inputenc}
 \usepackage{amsfonts}
 \usepackage{amsthm}
 \usepackage{amssymb, latexsym, amsmath}
 \usepackage{amsbsy}
\usepackage{amstext}
\usepackage{amsopn}
\usepackage{amsgen}
\usepackage[dvips]{color}
 \usepackage{graphicx}
  \usepackage{color}
\newcommand{\eb}{\begin{equation}}
\newcommand{\ee}{\end{equation}}
\shorttitle{Rotation of exoplanets and satellites}
\shortauthors{Makarov}
\begin{document}
\title{Equilibrium rotation of semiliquid exoplanets and satellites}
\author{Valeri V. Makarov}
\email{vvm@usno.navy.mil} 
\affil{US Naval Observatory, 3450 Massachusetts Ave NW, Washington DC 20392-5420, USA}

\date{Accepted . Received ; in original form }

\label{firstpage}
\begin{abstract}
A wide range of exoplanet and exomoon models are characterized by a finite average rigidity and a viscosity
much lower than the typical values for terrestrials. Such semiliquid bodies may or may not have
rigid crusts with permanent figures. Unlike planets with solid mantles and Earth-like rheology, semiliquid
bodies can be captured into stable pseudosynchronous spin resonance, where the average rate of rotation
is higher than the synchronous 1:1 resonance. Two basic conditions are derived for capture of planets
with a triaxial figure into pseudosynchronous
rotation, one related to the characteristic tidal wave number (the product of the tidal frequency by the Maxwell 
time), and the other to the orbital eccentricity. If
a semiliquid object does not satisfy either of the two conditions, it is captured into the synchronous resonance.
For nearly axially symmetric bodies, only the first condition is in place, and the other is much relaxed,
so they should predominantly be pseudosynchronous. It is also pointed out that the equilibrium pseudosychronous
rotation rate can not reach the widely used asymptotic value from the constant time lag model but is in reality
closer to the synchronous spin.
\end{abstract}


\section{Introduction}
Most of the Solar system bodies subject to significant tidal interaction are captured into the 1:1 spin-orbit
resonance, also known as synchronous rotation, when the body's average sidereal spin rate is exactly equal to the
orbital motion. As a result of this resonance, the body is permanently facing the perturber with the same side,
a phenomenon well known to us from the Moon. A notable exception to this rule is Mercury, which is captured
in a stable 3:2 resonance, which makes one Mercurian solar day as long as two Mercurian solar years. As was shown by
\citet{hut80}, synchronous rotation and a circular orbit is the terminal point of the tidal spin-orbit evolution;
however, higher stable spin-orbit resonances are possible and even likely for super-Earth exoplanets of presumably
terrestrial composition \citep{mabe}. Like Mercury, such planets, once caught in a super-synchronous resonance,
should stay there indefinitely long, unless an external impulse, such as a giant impact, drives the planet out of
this locally stable state. Most of the Solar system moons, on the other hand, are locked in the 1:1 resonance, and no
credible evidence was found of an equilibrium rotation with a non-resonant rate above the synchronous rate.

Investigation of equilibrium near-synchronous rotation states was pioneered by \citet{gol1} for the {\it ad hoc}
constant-$Q$ tidal model, in which the tidal torque is independent in magnitude of the perturbation frequency.
For a long time the work-horse of the tidal dynamics, the Constant Time Lag (CTL) model of tidal dissipation assumes
that the secular tidal torque acting on the perturbed body can be approximated by a linear dependence on tidal
frequency $\nu$ with a negative slope (analogous to a regular force of friction), but shifted upward for a non-vanishing
eccentricity \citep{mur}. As a result of the shift, the secular tidal torque is positive (accelerating in the prograde sense)
at the point of resonance, i.e., where the spin rate exactly equals the mean orbital motion, $\dot\theta=n$.
The secular term, which acts as friction in the balance of forces, vanishes at \citep{hut81,rod,fm08,we}
\eb
\dot\theta\simeq (1+6\,e^2)\,n
\ee

If the CTL model were applicable, in the absence of other forces, the moons of the Solar system would have been involved
in a slow prograde rotation relative to their host planets. The commonly accepted explanation why this does not happen\footnote{There are lingering doubts about the exact synchronism of Europa though \citep{gei}}
refers to the presence of a geometrical triaxial torque caused by the elongation of the permanent figure.
This torque emerges whenever the axis of the smallest inertia moment begins to deviate from the instantaneous direction
to the perturber; thus, it acts as a restorative force directed toward the point of exact synchronism.

Pseudo-synchronous equilibrium emerges in other, more advanced and complex tidal models as well. \citet{fm} considered a
rheological model dominated by Newtonian creep mechanism of dissipation and found an equilibrium tide at a super-synchronous
rate of rotation, although the shift turned out to be frequency-dependent in that case, as opposed to simply $6\,e^2 n$.
The hybrid viscoelastic model with self-gravitation and Andrade creep developed by \citet{efr1,efrb}, also permits
super-synchronous equilibria in principle. However, this state was shown to be practically impossible for bodies of terrestrial composition,
characterized by large effective viscosities and long Maxwell times \citep{me}, see also \citep{cor}.
On the other hand, a wide range of plausible rheological models for exoplanets and exomoons, as well as for smaller satellites
of the Solar system, can be best represented by the viscoelastic paradigm with low values of viscosity but finite rigidity.
The first important class of such objects is a terrestrial body that has been heated up by strong tidal forces past the
point of solidus in the deeper layers of the mantle.

\section{Torques}
\label{torq.sec}
In the following, we consider tides and forces acting on one of the components in a system of two orbiting bodies.
The component under consideration is called in this paper ``perturbed body", or ``rotating body", and in practice can be
a planet or a satellite. The other component is called ``perturbing body", or simply``perturber", and it can be
a star or a planet. The possible tides or deformations of the perturber are not involved in the analysis; it is
effectively a point-mass with only a central gravitational action on the perturbed body.

Comprehensive equations for the polar tidal torque (i.e., the torque component directed along the axis of rotation),
including the fast oscillating terms, can be found, for example, in \citep{mabe}. Here we only summarize the
main equations necessary for the following analysis of equilibrium rotation. 
The non-polar component of perturbation are ignored in this analysis, which is equivalent to setting
the orbit obliquity on the perturbed body's equator to zero.
The instantaneous torque acting on a rotating body is the sum of the triaxial torque caused by
the quadrupole inertial momentum, and the tidal torque caused by the dynamic deformation of
its body. Planets and moons of terrestrial composition have a permanent figure, which results
in the three principal moments of inertia $A$, $B$, and $C$ (the $C$ denoting traditionally the
largest moment of inertia) being unequal to one another. The degree of elongation of the permanent
figure, or the departure from the dynamical axial symmetry, can be represented by the parameter
$(B-A)/C$, which ranges from $\approx 0.1$ for smaller moons to $\approx 10^{-5}$ for the larger
rocky planets. The angular acceleration is related to the polar components of these torques by
\eb
\ddot\theta=\frac{T_{\rm TRI}+T_{\rm TIDE}}{\xi M_2R^2}
\label{eq.eq}
\ee
with $\theta$ being the rotation angle of the perturbed body reckoned from the axis of its largest elongation, corresponding to the moment $A$, $M_2$ is its mass, and $\xi$ being the coefficient of inertia, $C/(M_2R^2)$.
Possible variations of the moments of inertia on longer time scales are not considered in this paper.
The equations of the triaxial torque $T_{\rm TRI}$ have been exploited in the literature on many occasions
\citep{danb,gol,mur}. 
Neglecting
the precession and nutation of the planet, the angular acceleration due to the triaxial torque is
\eb
\ddot\theta_{\rm TRI}=-\frac{3}{2}\frac{B-A}{C}\frac{a^3}{r^3}n^2\sin2(\theta-f),
\label{tri.eq}
\ee
where $r$ is the instantaneous distance of the perturbed body from the perturber, $a$ is the semimajor axis,
$n$ is the orbital mean motion, and $f$ is the instantaneous true anomaly. This formula can be more
conveniently rewritten as an infinite series of mean anomaly (${\cal M}$) harmonics:
\eb
\ddot\theta_{\rm TRI}=-\frac{3}{2}\frac{B-A}{C}n^2\sum_j G_{20(j-2)}\sin(2\theta-j {\cal M}),
\label{kau.eq}
\ee
where $G_{lpq}$ are Kaula's functions of eccentricity related to Hansen's coefficients through $G_{20(j-2)}=X^{-3,\,2}_j$.
Note that the summation in Eq. \ref{kau.eq} is over $j=-\infty,\ldots,+\infty$, and $G_{20(-2)}=0$.

A simplified equation for the angular acceleration caused by tidal torque can be written as \citep[][Eqs. 106-108]{efr1}
\begin{eqnarray}
\ddot\theta_{\rm TIDE}&=&\frac{3}{2}\frac{M_1}{\xi M_2}n^2\left(\frac{R}{a}\right)^3\sum_{q=-1}^4 G_{20q}(e)
\sum_{j=-1}^4 G_{20j}(e) \nonumber\\  & &
\left[K_c(2,\nu_{220q})\,{\rm Sign}(\omega_{220q})\cos\left((q-j){\cal M}\right) -
K_s(2,\nu_{220q})\sin\left((q-j){\cal M}\right)\right],
\label{tide.eq}
\end{eqnarray}
assuming $M_2\ll M_1$ where $M_1$ is the mass of the perturbing body, and
\begin{eqnarray}
K_c(2,\nu_{220q})&\equiv& k_2(\nu_{220q})\sin\epsilon_2(\nu_{220q})\quad,\nonumber \\
\qquad K_s(2,\nu_{220q})&\equiv &k_2(\nu_{220q})\cos\epsilon_2(\nu_{220q})\quad,
\end{eqnarray}
$k_2$ being being the degree-2 dynamical Love number, $\epsilon_2(\nu_{220q})$ the positive definite, mode-dependent 
tidal phase lag and $\nu_{220q}$ being the positive definite
physical forcing frequency related to the semidiurnal Fourier mode through
\eb
\nu_{220q}=|\omega_{220q}|=|(2+q)n-2\dot\theta|.
\ee
One of the functions, $K_c$, has been scrutinized in the literature \citep{efr1,efrb}. Historically, it is denoted
as $k_2/Q$, also dubbed kvalitet elsewhere, the $Q$ value being related to the amount of energy dissipated over one cycle of rotation. Secular
(i.e., non-averaging over one orbital period) acceleration emerges from the $K_c$ term when $q=j$. 

Within the simplifying assumption of a spherical, incompressible, and homogeneous body, and a purely analytical
approach adopted in this paper, the corresponding kvalitet can be written as \citep{efr2,efr1,efrb}
\eb
K_c(2,\nu)=-\frac{3}{2}\frac{{\cal A}_2\,\nu\tau_M \Im(\nu)}{(\Re(\nu)+{\cal A}_2\nu\tau_M)^2+\Im^2(\nu)}
\label{qual.eq}
\ee
where
\eb
{\cal A}_2=4\pi \frac{19}{6}\frac{ R^4\mu}{{\cal G}M_2^2},
\ee
${\cal G}$ being the gravitational constant, and
\begin{eqnarray}
\Re(\nu)&=&\nu\tau_M+\nu^{1-\alpha}\tau_M\tau_A^{-\alpha}\cos\left(\frac{\alpha\pi}{2}\right)\Gamma(1+\alpha)
\nonumber\\
\Im(\nu)&=&-1-\nu^{1-\alpha}\tau_M\tau_A^{-\alpha}\sin\left(\frac{\alpha\pi}{2}\right)\Gamma(1+\alpha)
\label{compl.eq}
\end{eqnarray}
The $\Re(\nu)$ and $\Im(\nu)$ dimensionless functions are the real and imaginary parts of the 
normalized dimensionless complex compliance,
respectively, multiplied by tidal wave number $\nu\tau_M$ to avoid numerical singularity at zero frequency. In these equations, $R$ is the radius of the perturbed body, $M_2$ is its mass, 
$\tau_M$ is the Maxwell time, $\tau_A$ is the Andrade time, $\mu$ is the unrelaxed shear modulus. The frequency
$\nu$ is used hereafter as a shorthand version of $\nu_{220q}$, so the functions remain dependent on
both the frequency and the tidal mode. The dimensionless
Andrade parameter $\alpha$ takes values from $\sim 0.16$ for a partial melt to $\sim 0.4$ for cold minerals.
The parameter ${\cal A}$, 
sometimes called effective rigidity in
the literature, is related to the definition of the static Love number \citep{gol}.
The second terms of the complex compliance components represent the contribution of the Andrade creep mechanism.

\begin{figure}[htbp]
\epsscale{1.15}
  \plottwo{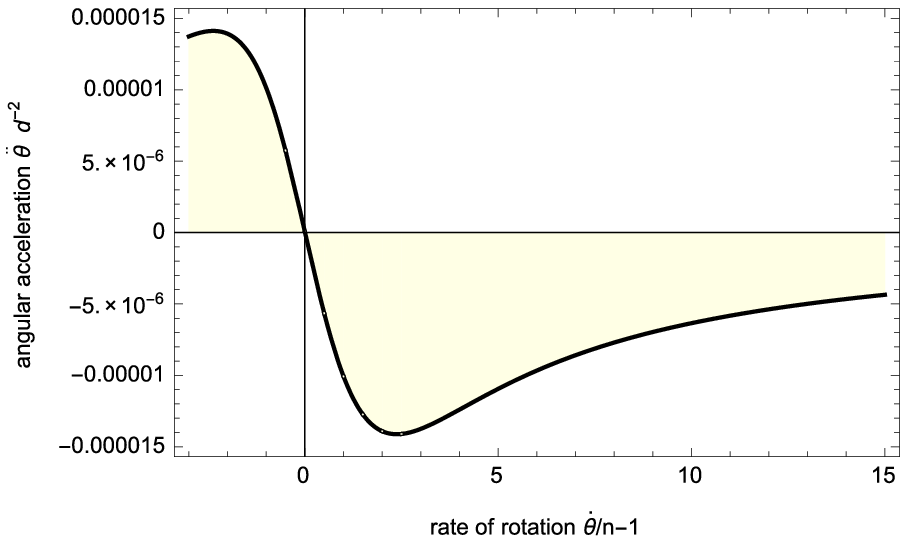}{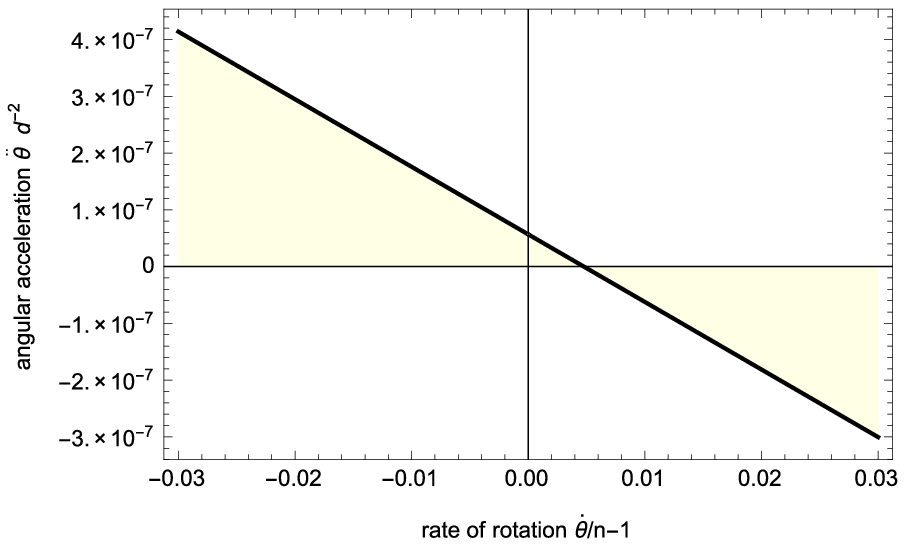}
\hspace{2pc}
\caption{Angular acceleration of Titan caused by the secular component of tidal torque (Eq. \ref{tide.eq})
with the assumed parameters
listed in Table \ref{param.tab} and with a Maxwell time of 0.1 days. Left: a large-scale view of the dependence showing the
main ``kink" corresponding to the 1:1 resonance term, $\omega_{2200}$. Right: the same dependence on much smaller scale to
demonstrate the offset of null torque from the exact 1:1 resonance. The kinks associated with the other resonances
$\dot\theta/n=0.5,1.5,2.0,2.5,\ldots$ are too small in amplitude and smoothed out to be discerned.}
\label{kvalitet.fig}
\end{figure}

These basic equations greatly simplify for the case of semiliquid bodies considered in this paper. Recalling that
the Maxwell time is defined as $\tau_M=\eta/\mu$, where $\eta$ is the effective viscosity, and that the Andrade
creep is perhaps insignificant for such bodies, one can use the constraints $\tau_A=\infty$ and $\tau_M\,n\ll 1$.
These simplify the secular kvalitet to
\eb
K_c(2,\nu) \equiv k_2(\nu)~\sin\epsilon_2(\nu)=\frac{3}{2}\frac{{\cal A}_2\lambda}{\lambda^2(1+{\cal A}_2)^2+1},
\label{k.eq}
\ee
where $\lambda=\tau_M\nu$ is a dimensionless quantity hereafter termed the tidal wave number. 
This function is strongly dependent on tidal frequency in narrow
vicinities of spin-orbit resonances $(2+q)n=2\dot\theta$ for integer $q$. 
The characteristic kink-shape of the near-resonant torque has the largest amplitude in the 1:1 resonance for small
and moderate orbital eccentricities. 
A steep decline
between the two peaks (Fig. \ref{kvalitet.fig}) occupies a narrow band of frequencies for realistic rheologies. Despite the relatively
small amplitude of the kink (compared to a typical amplitude of the triaxiality-caused torque), it acts
as an efficient trap for a planet trying to traverse the resonance. The maximum kvalitet is reached at
\eb
\lambda_{\rm max}=\frac{1}{1+{\cal A}_2},
\label{max.eq}
\ee
where its value is 
\eb
K_c(2,\nu_{\rm max})=\frac{3}{4}\frac{{\cal A}_2}{1+{\cal A}_2}.
\ee

\section{First condition of pseudosynchronous equilibrium}
\label{first.sec}
As is evident from Eq. \ref{max.eq}, the locale of maximum tidal force in frequency is inversely proportional to
the effective Maxwell time. Relatively cold terrestrial mantles are characterized by large values of viscosity
\citep[e.g.,][]{dor}, and, thus, long Maxwell times. For example, the tidal response of Mercury was modeled with
a $\tau_M=500$ years \citep{noe}, and the suggested value for the Moon is $\sim 10$ years as estimated from the
observed tilt of the mean principal axis \citep{mamoon}. Icy satellites, on the other hand, can have much shorter
Maxwell times owing to the relatively low viscosity of pressurized ice ($\sim 10^{15}$ Pa$\cdot$ s). The slope
of the central part of the secular torque between the opposing peaks (Fig. \ref{kvalitet.fig}, left), 
which can be roughly approximated with an inclined line, is there fore proportional to the Maxwell time.
Within this segment of the kink-shape, the secular torque acts as a quasi-linear friction force
on any deviations of the spin rate from the equilibrium value. The point of equilibrium, or zero secular torque,
is associated with the root of the torque curve, which is offset from the point of synchronous resonance
because of the positive, nearly constant, component due to the second-largest term $q=1$ for a finite eccentricity
(Fig. \ref{kvalitet.fig}, right). The tidal torque drives the rotating body away from the point of synchronism
$\nu_{2200}=0$ toward the point of equilibrium $\nu_{2200}=\nu_{\rm pseudo}$ when the frequency takes values within
the narrow segment between these points. Obviously, the point of pseudosynchronous equilibrium, where
the secular torque vanishes, can not be outside of the linear segment bracketed by the opposite peaks.
Therefore, the asymptotic value of $(1+6\,e^2)n$ can only be realized when this frequency is much smaller
than the location of the peak $\tau_M^{-1}\,(1+{\cal A}_2)^{-1}$, i.e., in liquid or semiliquid
bodies. For inviscid planets with Maxwell time comparable to, or greater than, the orbital period of the perturber,
the point of equilibrium collapses to very small values. This practically eliminates the feasibility of
pseudosynchronous equilibrium, as any small perturbation, including physical libration, will drive the body
into the globally stable synchronous rotation.

\begin{figure*}
\begin{minipage}{82mm}
\includegraphics[width=82mm]{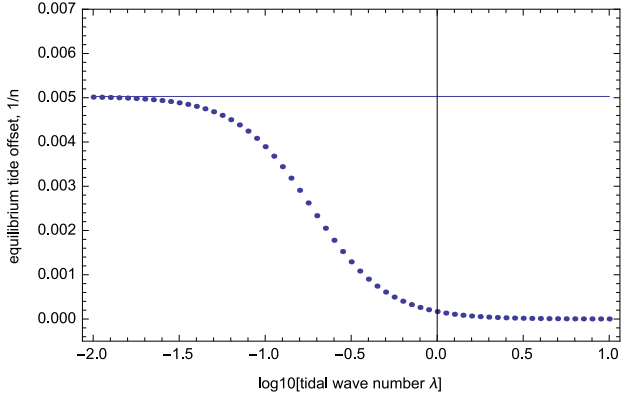}
\end{minipage}\hspace{2pc}
\caption{Normalized frequency of equilibrium, i.e. of vanishing secular tidal torque, as a function of tidal
wave number $\lambda$. The asymptotic value $6\,e^2$ is shown with the horizontal line.}
\label{pseudo.fig}
\end{figure*}

Fig. \ref{pseudo.fig} shows the relative, normalized frequency of pseudosynchronous equilibrium as a function of tidal wave
number for the Titan model (Table \ref{param.tab}). This graph is obtained by computing the roots of
Eq. \ref{tide.eq} for the secular part of tidal acceleration ($q=j$) and for different values of $\tau_M$.
The previously known point of $(1+6\,e^2)\,n$ from the CTL formalism, indicated by the horizontal line in the graph,
is only an asymptotic limit achieved at $\lambda \rightarrow 0$. Realistic viscoelastic bodies with self-gravitation
attain the pseudosynchronous equilibrium at any rotation rate between $n$ and $(1+6\,e^2)\,n$ depending on the effective
wave number $\tau_M\,n$.

 \begin{table*}
 \centering
 \caption{Parameters of the tidal model for Titan as a tentative semiliquid satellite.}
 \label{param.tab}
 \begin{tabular}{@{}lrrr@{}}
 \hline
            &                 &       &\\
   Name     &  Description    & Units & Values\\
 \hline
 $\xi$ & \dotfill moment of inertia coefficient &   &  0.34\\
 $R$   & \dotfill radius of planet              & m & $2.576\cdot 10^6$\\
 $M_2$ & \dotfill mass of the perturbed body (satellite)& kg & $1.3452\cdot 10^{23}$\\
 $M_1$ & \dotfill mass of the perturbing body (planet)& kg & $5.68\cdot 10^{26}$ \\
 $a$ & \dotfill semimajor axis & m & $1.2219\cdot 10^{9}$\\
 $n$ & \dotfill mean motion, i.e. $2\pi/P_{\rm orb}$ & d$^{-1}$ & 0.394\\
 $e$ & \dotfill orbital eccentricity & & 0.0288\\
 $(B-A)/C$ & \dotfill triaxiality & & $1.3\cdot 10^{-4}$ \\
 $\cal{G}$ & \dotfill gravitational constant & m$^3$ kg$^{-1}$ yr$^{-2}$ & $66468$\\
 $\tau_M$ & \dotfill Maxwell time  & yr &  var.\\
 $\tau_A$ & \dotfill Andrade time  & yr &  $\infty$\\
 $\mu$ & \dotfill unrelaxed rigidity modulus & Pa
  & $3\cdot10^{9}$\\
 $\alpha$ & \dotfill the Andrade parameter & & $0.2$\\
 \hline
 \label{table}
 \end{tabular}
 \end{table*}

As is shown in Fig. \ref{pseudo.fig}, the point of equilibrium becomes close to the CTL limit at wave numbers as small
as 0.1. This implies a Maxwell time of 10 days or shorter for the Titan model (Table \ref{param.tab}). The fact that Titan
is in synchronous rather than pseudosynchronous rotation can be interpreted as Titan's effective Maxwell time being
longer than that value. However, other circumstances of purely dynamical, rather than rheological, character can
prevent the capture of inviscid bodies into pseudosynchronism.

\section{Second condition of pseudosynchronous equilibrium}
\label{second.sec}

Capture into spin-orbit resonances, including the 1:1 (synchronous) resonance, is a probabilistic process, conditioned
by the balance of restorative and damping forces similar to a nonlinear, driven rotating pendulum with friction \citep{gold,
gol}. The subtle conditions of a {\it passage} through spin-orbit resonances were numerically explored by \citep{ma12}.
In general, at each resonance $\dot\theta/n=$ 1:2, 1:1, 3:2, 2:1, $\ldots$, the resonant term of secular torque
will be biased (or offset from
zero) by the contribution of the other secular ($q=j$) terms. For example, the bias at the 1:1 resonance in secular torque is
positive (see Fig. \ref{kvalitet.fig}, right), 
mostly defined by the left-hand shoulder of the 3:2 secular term, for moderate eccentricities. Physically, the positive
bias is generated by the fact that the perturber moves with a higher than $n$ angular velocity during one half of the
orbital period, and the bodies are closer to each other during that part of the orbit than the other half. The
emerging positive, or accelerating in the prograde sense torque, tends to spin up the synchronized perturbed body
\citep{mur}. On the
contrary, the bias at the 3:2 resonance is usually negative (decelerating) at moderate eccentricities, mostly defined
by the right-hand shoulder of the 1:1 term. It is in fact this bias that generates the possibility of non-resonant
equilibrium, or pseudosynchronism. The bias can well be greater in magnitude than the half-amplitude of a resonant
kink, in which case there is no equilibrium of secular torque at all (however, capture into that resonance may still be possible). Obviously, the bias acts as a driving force, trying to drag the rotating body through a resonance,
whereas the near-linear inclined segment between the peaks acts as a damping force, or friction, working to get rid of the
excess kinetic energy and to capture the body into the resonance. 

\begin{figure}[htbp]
\epsscale{1.15}
  \plottwo{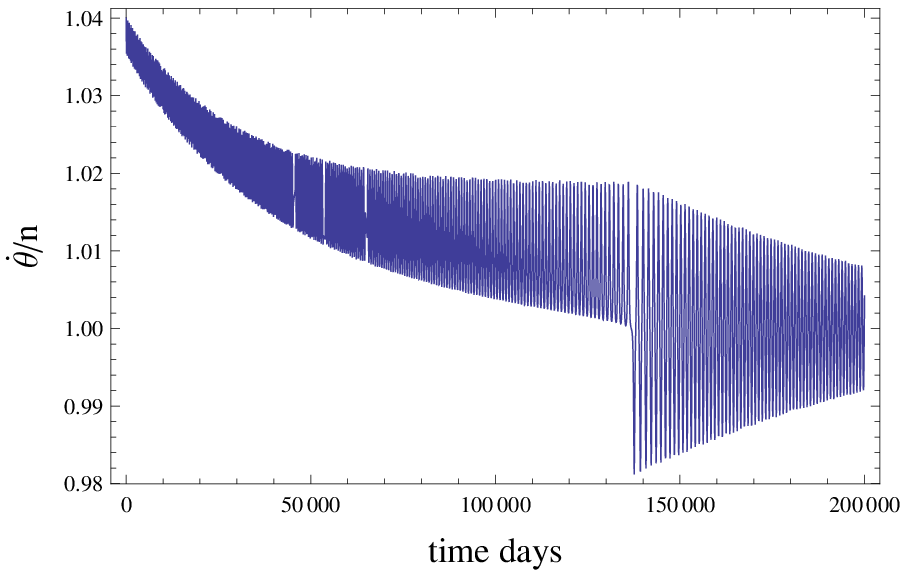}{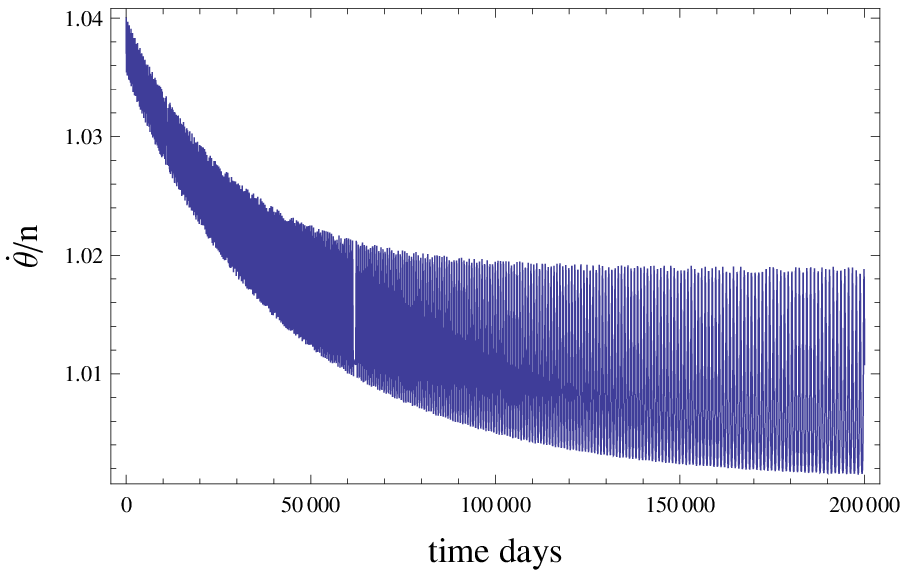}
\hspace{2pc}
\caption{Numerical integration of Titan's spin rate approaching the 1:1 spin-orbit resonance for two different values
of orbital eccentricity. Left: capture into synchronous rotation with $e=0.045$. Right: capture into pseudosynchronous
rotation with $e=0.046$. }
\label{capture.fig}
\end{figure}

Returning to the useful analogy with a driven rotating pendulum, the mechanics of a resonance encounter can be
visualized as a gradual slowing of rotation due to the secular tidal torque. With each cycle, the minimum angular
velocity at the top of the circular path becomes smaller, until the pendulum stops close to the top point,
no longer able to reach it. It reverses the motion, suddenly switching from circulation to oscillation. The first
full-amplitude swing in the counter direction determines the fate of the pendulum. If the work done by the friction force 
is greater than the work done by the driving torque, the pendulum will not be able to come over the top in the opposite
direction either, and it will continue to oscillate around the point of equilibrium with an ever decreasing amplitude
due to the damping. In the numerical simulation shown in Fig. \ref{capture.fig} (left), this capture into the
synchronous resonance takes place at about 137000 days. 
On the other hand, when the driving torque is strong relative to the friction, the pendulum
will pass over the top and continue to circulate in the opposite direction.

Capture into a pseudosynchronous equilibrium state may occur when the secular tidal torque averaged over one circulation
period, vanishes due to the offset in relative frequency $\dot\theta/n-1$ (Fig. \ref{kvalitet.fig}, right)
while the minimum velocity at the top of the trajectory
is still greater than the resonant value. In other words, the pendulum still rotates in the prograde direction
but the secular torque is balanced out. This situation is depicted in Fig. \ref{capture.fig} (right) for the same set
of model parameters. The mean velocity $\dot\theta$ remains above $1\,n$ indefinitely long, because the lowest
velocity over a free libration cycle can not reach the resonant value. Should the resonant value be reached,
an immediate transition into the synchronous state happens.

The emerging condition of pseudosynchronous equilibrium can be quantified using the pre-resonant trajectory for
free librations. Let us introduce, following \citep{gol},
\eb
\gamma=2\theta-2\cal M,
\label{gamma.eq}\ee
which is the sidereal libration angle multiplied by 2. Large-amplitude free librations of bodies with a permanent
figure are driven by the
triaxial torque (Eq. \ref{tri.eq}). For the case of small eccentricity, which is valid for the majority of
close exoplanets detected so far, as well as for the majority of Solar system satellites, only the dominating
term from Eq. \ref{kau.eq} can be retained leading to
\eb
\ddot\gamma\simeq -3 \frac{B-A}{C} n^2 G_{200}(e) \sin\gamma.
\ee
Note that this approximation can only be used for $e<0.1$. From the consideration of energy conservation
\eb
\dot\gamma^2=\dot\gamma_0^2+2\epsilon^2n^2(1+\cos\gamma),
\label{kin.eq}
\ee
where $\epsilon=\sqrt{3 G_{200}(e)(B-A)/C}$, and $\dot\gamma_0$ is the minimum libration velocity at $\gamma=\pi$. 
Equilibrium is achieved when the kinetic energy averaged over one libration period is constant, which requires
the work done by the secular tidal torque to be zero, i.e.,
\eb
\int_0^{2\pi}\ddot\theta_{\rm TIDE}\;d\gamma=0.
\label{work.eq}
\ee
Per condition of pseudosynchronism number one
(Section \ref{first.sec}), the tidal wave number of the resonant mode should be quite small, in which case the
kvalitet in Eq. \ref{k.eq} becomes a linear function of $\nu$
\eb
K_c(2,\nu)\simeq\frac{3}{2}{\cal A}_2\lambda,
\ee
and the secular tidal acceleration can be approximated quite well by a linear function,
\eb
\ddot\theta_{\rm TIDE}\propto -(\dot\gamma-\dot\gamma_{\rm pseudo}).
\label{linear.eq}
\ee
Combining Eqs. \ref{kin.eq}, \ref{work.eq}, and \ref{linear.eq} obtains
\begin{eqnarray}
\dot\gamma_{\rm pseudo}&=&\frac{1}{2\pi}\int_0^{2\pi}\left(\dot\gamma_0^2+2\epsilon^2n^2(1+\cos\gamma)\right)^{1/2}\;d\gamma
\nonumber \\
 &=&\frac{2}{\pi}\left[
\sqrt{\dot\gamma_0^2+4\epsilon^2n^2}\:{\cal E}\left(\frac{4\epsilon^2n^2}{\dot\gamma_0^2+4\epsilon^2n^2}\right)\right]
\end{eqnarray}
where ${\cal E}$ stands for the complete elliptic integral. Because of the character of the secular tidal torque (Fig. \ref{kvalitet.fig},
right), the planet will decelerate if the relative root-mean-square (RMS) velocity over one libration period is higher 
than $\dot\gamma_{\rm pseudo}$. By the same argument, the planet will accelerate if its RMS velocity is below the point
of pseudosynchronism, unless it is captured into the 1:1 resonance.  To avoid the resonance, 
the minimum velocity should be positive,
$\dot\gamma_0>0$. In the marginal case $\dot\gamma_0=0$, the period of free libration becomes infinitely long,
because the pendulum stops in the unstable equilibrium $\ddot\gamma=\dot\gamma=0$ and $\gamma=\pi$. The time-average spin rate,
therefore, nullifies. Remarkably, the time-average kinetic energy or the RMS velocity do not nullify, in that as the
minimum velocity approaches 0, $\dot\gamma_0\rightarrow 0$, the RMS velocity approaches a finite positive number, namely,
\eb
\dot\gamma_{\rm pseudo}\rightarrow \frac{4}{\pi}\:\epsilon\: n.
\ee
Introduce a nominal eccentricity $e_{\rm crit}$ such that the secular component of tidal torque vanishes at a
frequency $6\,e_{\rm crit}^2\,n$. As discussed in Section \ref{first.sec}, $0<e_{\rm crit}<e$. By this definition, the 
relative spin rate $\dot\theta_{\rm pseudo}-n$ at
pseudosynchronism should be equal to $6\,e_{\rm crit}^2\,n$, where $e_{\rm crit}\rightarrow e$ asymptotically for
$\lambda \rightarrow 0$. Taking into account Eq. \ref{gamma.eq}, we finally arrive at the second
necessary condition of pseudosynchronus equilibrium:
\eb
e > \left(\frac{\sqrt{B-A}}{\sqrt{3C}\:\pi}\right)^{1/2}.
\label{ecrit.eq}
\ee
A similar condition, only different by a constant factor $\sqrt{19}/2$, was derived by \citet{gol1} for the rough
constant-$Q$ tidal model.

\section{Calculations for a model of Titan}
\label{titan.sec}

Titan is one of the largest satellites in the Solar system, and a good representation of the class of icy moons
\citep{hus}. Objects like Titan can be common in exoplanet systems at relatively larger separations from their
hosts, bridging the gap between ice-dominated and terrestrial (silicate) objects. A series of numerical integrations
was performed for this study within the uniform body approximation for a set of parameters listed in Table 1.
The degree of triaxiality $(B-A)/C$ was deduced from the numbers provided by \citet{bil}. The Andrade time was set
equal to infinity, thus nullifying the effects of inelastic creep of the Andrade type. For most simulations,
the Maxwell time was assumed to be a rather short 0.1 days, which, for an effective shear modulus of $3\cdot 10^9$ Pa
(suggested by F. Nimmo, priv. comm. 2013), implies a rather low effective viscosity of the dissipating layer of
$2.6\cdot 10^{13}$ Pa$\cdot$s. This may still be a plausible value for a loosely bound mixture of rock and ice,
as well as partially molten mantles. A short Maxwell time is required to realize the pseudosynchronous equilibrium
according to the first condition (Section \ref{first.sec}), i.e., $\tau_M\,\nu \la 0.1$. The asymptotic RMS
velocity of a pseudosynchronously rotating body is $\dot\theta_{\rm pseudo}=(1+6e^2)\,n$, hence the corresponding
tidal frequency in the main mode is $\nu=12\,e^2\,n$, which amounts to 0.004 d$^{-1}$ for the present day eccentricity,
and to 0.009 d$^{-1}$ for $e_{\rm crit}$ (Section \ref{second.sec}). Therefore, for the Titan model in question,
the first condition of pseudosynchronism is fulfilled for Maxwell times shorter or approximately equal to 10 days.
We note that the first condition can be re-written for the
effective viscosity to be
\eb
\eta \la \frac{0.1\,\mu}{12\,e^2\,n}.
\label{eta.eq}
\ee
Interestingly, for a fixed finite viscosity, the eccentricity can not be very large lest the pseudosynchronism 
equilibrium collapses by the first condition. Therefore, there is a certain interval of effective viscosity
for pseudosynchronism to work. This interval vanishes at a characteristic viscosity
\eb
\eta_{\rm crit}=\frac{0.1\sqrt{3}\,\pi\,\mu}{12\,\sqrt{(B-A)/C} n}\approx 0.045 \frac{\mu}{\sqrt{(B-A)/C} n},
\label{eta.eq}
\ee
so that for larger values of viscosity, this type of equilibrium is practically impossible. For our Titan
model, the critical viscosity turns out to be $2.6\cdot 10^{15}$ Pa$\cdot$s. This is much smaller than the typical
effective viscosity of cold terrestrial mantles ($\eta_{\rm Earth}\approx 1.3\cdot 10^{21}$ Pa$\cdot$s). On the
other hand, due to the Arrhenius-type dependence of viscosity on temperature \citep[e.g.,][]{dor}, the average
viscosity drops to $\simeq 1.3\cdot 10^{15}$ Pa$\cdot$s at the bottom of the Earth's mantle (temperature 4000 K,
pressure 136 GPa) close to the liquidus, according to both the models presented by \citet{sho} and \citet{sta}.
Similarly small pre-solidus viscosity is reached in the models considered by \citet{hen}.
With a fixed $\mu$, planets and satellites with smaller triaxial deformation or more distant from their hosts
have more room for a locally stable non-synchronous state, in accordance with Eq. \ref{eta.eq}. The triaxial
deformation of smaller Saturnian satellites is so large \citep{tis}, that it probably eliminates the possibility
of pseudosynchronism.

Several numerical simulations of spin evolution for the Titan model were performed to verify the second condition
of pseudosynchronism (Section \ref{second.sec}). The input parameters were kept as in Table \ref{param.tab} except
for the eccentricity, which varied between 0.02 and 0.06. Each integration solves numerically an ordinary differential
equation of second order (Eq. \ref{eq.eq}) for the angular acceleration including both the triaxial (Eq. \ref{tri.eq})
and tidal (Eq. \ref{tide.eq}) torques. The initial conditions were set to $\theta(0)=0$, $\dot\theta(0)=1.4\,n$,
so that the moon would start far enough from and above the 1:1 resonance to allow for a free, dissipative descent
into an equilibrium. The standard Runge-Kutta integration algorithm in Mathematica was employed with an integration step
of 0.1 days and an accuracy goal of 9 significant digits. The tidal torque included both the secular and oscillating
terms (although the latter are probably negligible in the presence of much larger free librations), and the summation of
tidal modes was performed over indices $j=-1,0,\ldots,4$, as per Eq. \ref{tide.eq}, which is quite sufficient for
such small eccentricities. The length of integration was set at $200000$ days, which would assure a capture into one
or the other equilibrium. 

Fig. \ref{capture.fig} shows the resulting relative spin rate for two fixed values of eccentricity, $e=0.045$ (left)
and $e=0.046$ (right). This small change in the orbital eccentricity results in a dramatically different outcome of
the encounter with the dominating resonance. In the former case, the spin rate at the minimum of a free libration
cycle reaches the resonance $1\,n$, followed by immediate capture into that resonance. In the latter case, the
point of equilibrium torque rotation is far enough from $1\,n$, the minimum spin rate can never quite reach the
resonance, and the body remains in a super-synchronous rotation. According to Eq. \ref{ecrit.eq}, the theoretical
threshold eccentricity enabling pseudosynchronism is 0.0449. These computations show that the actual threshold
eccentricity is between 0.045 and 0.046. The cause for this slight difference is in the asymptotic character of the
assumed equilibrium velocity $(1+6\,e^2)n$ (Fig. \ref{pseudo.fig}).

\section{Escapes from equilibrium}
\begin{figure*}
\begin{minipage}{82mm}
\includegraphics[width=82mm]{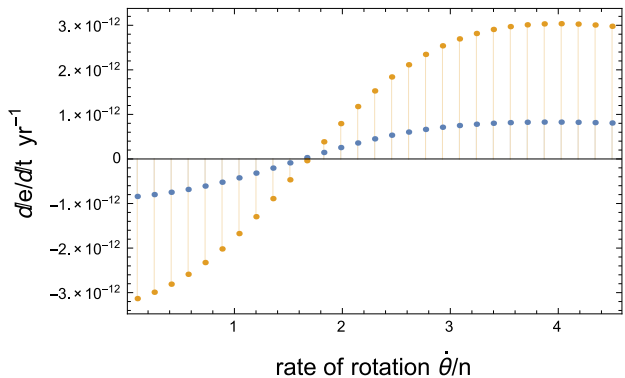}
\end{minipage}\hspace{2pc}
\caption{Rate of secular eccentricity change for the Titan model (Table 1) with two fixed values of
eccentricity: $e=0.1$ (outer curve) and $e=0.0288$, or the current value (inner curve).}
\label{dedt.fig}
\end{figure*}

Our considerations have been thus far decoupled from the orbital evolution processes caused by the secular
components of the polar tidal torque. A comprehensive analysis of the problem should, in principle,
include the total energy and angular momentum of the system compounded of the rotational and orbital motion of
the bodies. The classic Kaula's approach, on the other hand, is based on the non-inertial reference frame
related to the equator of the perturbed body, and is not amenable to a reformulation in a barycentric
non-rotating reference frame. Some justification to this approach is provided by the long-known fact that
the characteristic time of spin rate evolution is typically much shorter than the time of orbit evolution
driven by tides \citep{zah,auc}. On the time scales of the order of $10^5$ years (Fig. \ref{capture.fig}), it is
completely legitimate to ignore the secular changes in the orbital parameters. On the time-scales of
a stellar age, however, these changes should be taken into account.

The orbital eccentricity, in particular, is a crucial parameter involved in the conditions of pseudosynchronism. Using
Kaula's frequency decomposition of the perturbing potential approach \citep{kau,lam,rem},
we have for the rate of eccentricity from the dominating $l=2$ terms\footnote{For comparison, see similar
equation derived for the CTL model in \citep{mig80,mig81}.}
\begin{eqnarray}
&\langle \frac{de}{dt}\rangle = 
-\frac{n\sqrt{1-e^2}}{(M_1+M_2)a^5\,e}\; \sum_{m=0}^2 \frac{(2-m)!}{(2+m)!}(2-\delta_{0m})  \\
\times&\sum_{p=0}^2 \sum_{q=-\infty}^{+\infty}G_{2pq}^2(e)\,\left[\sqrt{1-e^2}(2-2p+q)-(2-2p)\right] \nonumber\\
\times&\left[M_2 R_1^5\,F_{2mp}^2(i_1)K_1(2,\nu_{2mpq}^{(1)})\,{\rm Sign}(\omega_{2mpq}^{(1)})
 +
M_1 R_2^5\,F_{2mp}^2(i_2)K_2(2,\nu_{2mpq}^{(2)})\,{\rm Sign}(\omega_{2mpq}^{(2)})\right]
\nonumber\label{dedt.eq}
\end{eqnarray}
where $K_1$ and $K_2$ are the tidal quality functions from Eq. \ref{qual.eq} for the primary and the secondary
components (Saturn and Titan, in our case), respectively, $i_1$ and $i_2$ are the obliquities of the orbital
plane on the respective equators, and $R_1$ and $R_2$ are the radii. Ignoring the term corresponding to
the dissipation in Saturn, which is probably small\footnote{See, however, a small $Q$ for Saturn in
\citep{lai}  estimated from astrometric observations of its satellites.}, this equation implies a 
negative rate of eccentricity
everywhere in the neighborhood of the 1:1 resonance for a range of past eccentricities, see Fig. \ref{dedt.fig}. 
Thus, a Titan in pseudosynchronous or synchronous equilibrium should be slowly damping the eccentricity, 
barring third-body interactions. The upper bound equilibrium frequency, being proportional to $e^2$,
will decline faster, squeezing the range of pseudosynchronism. Eventually, the critical eccentricity
is achieved and the body falls into the 1:1 resonance.
\begin{figure*}
\begin{minipage}{82mm}
\includegraphics[width=82mm]{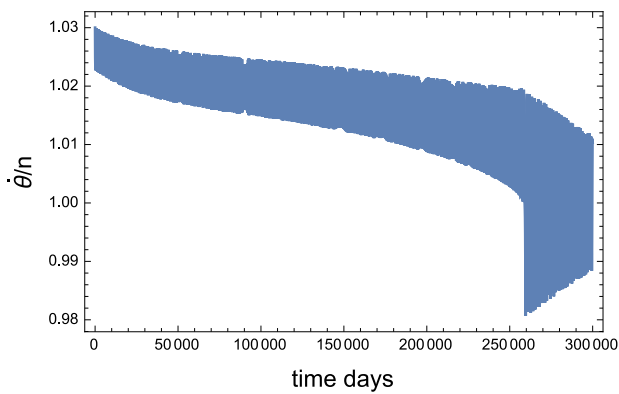}
\end{minipage}\hspace{2pc}
\caption{Numerically simulated transition of Titan's spin rate from pseudosynchronous equilibrium to the 1:1
(synchronous) resonance caused by a gradual decline of orbital eccentricity from 0.06 at $t=0$ to $0.042$ at
$t=300000$ days.}
\label{transit.fig}
\end{figure*}

This process was confirmed by numerical simulations. Fig. \ref{transit.fig} depicts the results of a numerical
simulation similar to the ones in Fig. \ref{capture.fig} but for a longer time, and with a variable eccentricity.
The initial parameters were set to $\theta(0)=0$, $\dot\theta(0)=1.03\,n$, while a slowly declining eccentricity
was modeled as $0.06\times \cos(3\cdot 10^{-6}\,t)$, bringing it down to approximately 0.04 by the end of the
integration. If the initial eccentricity were kept fixed during the integration span, the model body would
descent into the pseudosynchronous state and remain there indefinitely long staying clear of the resonance.
But the declining eccentricity pushed the RMS velocity down, the minimum free libration velocity reached
$1\,n$, and immediate capture into the 1:1 resonance followed. 

\section{Rotationally symmetric planets}
Thus far, we have considered celestial bodies which have nonzero elongation in the distribution of mass and in shape,
when in the sequence of principal moments of inertia $A\le B\le C$, the smallest two are unequal, $B>A$. This class
of objects is of special practical interest, as it presumably includes most Earth-sized planets and smaller
satellites. Only the greater planets of the Solar system do not fall into this category. To a first-order approximation,
moons with deep oceans covered by a rigid surface crust can also be described by our model, as long as the crust
is not perfectly symmetric and is intact as a whole.\footnote{A more accurate and consistent treatment of bodies
with subsurface oceans requires more complex modeling including tidal wave propagation and boundary friction, best
handled by numerical methods.} Other plausible configurations include planets with massive surface oceans but
rigid cores, the latter likely to have nonzero triaxial deformation. Finally, close-in terrestrial planets can be molten
to the surface by tidal dissipation or stellar irradiation, in which case they can not maintain a permanent figure.
A bulge can only be raised by external perturbations, and should be dynamic and transient in nature. It is not
obvious if the viscoelastic model in question is at all valid for such bodies. However, the commonly employed CTL model
is mathematically similar to the Maxwell model in the linear regime. Therefore, the case of $A=B$ may at least be of
theoretical interest, delineating the boundaries of our knowledge within the existing analytical models.

Obviously, the second condition of pseudosynchronism (Eq. \ref{ecrit.eq}) is automatically satisfied. At first glance,
the first condition also appears defunct, because any finite separation of the point of equilibrium secular torque
from exact resonance should guarantee capture into this equilibrium in the absence of oscillations. Indeed, there is no
triaxial torque, thus, there is no free libration. A more careful analysis, supported by numerical experiments,
shows that other, much smaller librations are still present. These are always harmonics of the orbital frequency
caused by the periodic terms of the tidal torque, Eq. \ref{tide.eq}. We observe that the mixed terms $q\neq j$ give
both sine- and cosine-harmonics of perturbing polar torque, and hence, similar variations of the spin rate and libration
angle. For a given pair $q,j$, the difference between these harmonics is in the kvalitet functions $K_c$ and $K_s$.
Recalling that $G_{20(-1)}=-1/2\;e+O(e^3)$, $G_{200}=1-5/2\;e^2+O(e^4)$, $G_{201}=7/2\;e+O(e^3)$, and all other
Kaula functions are $O(e^2)$, or higher orders in $e$, our analysis can be safely limited to the combinations of
0 with -1 or 1 in the $(q,j)$ indices, for a small eccentricity. Furthermore, as long as the spin rate is close
to the 1:1 resonance, the largest by magnitude terms among the cos-harmonics will be the ones with $q=\pm 1$ and
$j=0$, because the $q=0$ term has $K_c(2,0)=0$. The kvalitet for these two terms is the same, equal to $K_c(2,n)$ (Eq. \ref{k.eq}).
For the case $M_2 \ll M_1$, this leads to an approximate equation for the combined cos-harmonic of tidal forced libration
\eb
\ddot\theta_{\rm cos}\approx 6\frac{M_1}{\xi M_2}n^2\left(\frac{R}{a}\right)^3 K_c(2,n)\,e\,\cos{\cal M}.
\ee
The corresponding forced libration in angular velocity is $\dot\theta-n\propto \sin{\cal M}$, and in angle $\theta-{\cal M}
\propto -\cos{\cal M}$. Apart from being numerically very small, this perturbation can not violate the pseudosynchronous
equilibrium, because it is out of phase with the orbital acceleration of the perturber. Indeed, in the 
absence of a triaxial force,
any restorative force should still be proportional to $+\sin{\cal M}$ to lock the body into synchronous rotation. 

The sin-harmonics of oscillating tidal torque emerge in the general Darwin-Kaula expansion \citep{efr1}, but,
to my knowledge, have not been exploited in the literature. They are proportional to the kvalitet functions
$K_s(2,\nu)$ (Section \ref{torq.sec}, Eq. \ref{tide.eq}), which are, explicitly,
\eb
K_s(2,\nu)=\frac{3}{2}\frac{{\cal A}_2\nu\tau_M\Re(\nu)+ \Re(\nu)^2+\Im(\nu)^2}{(\Re(\nu)+{\cal A}_2\nu\tau_M)^2+\Im^2(\nu)}.
\ee
For negligible Andrade creep contribution ($\tau_A=\infty$), this further simplifies to
\eb
K_s(2,\nu)=\frac{3}{2}\frac{{\cal A}_2\lambda^2+ \lambda^2+1}{\lambda^2(1+{\cal A}_2)^2+1}.
\ee
where $\lambda=\nu\tau_M$ is the tidal wave number. The maximum value of $K_s$ is reached at $\nu=0$, $K_s(2,0)=3/2$,
whereas $K_c(2,\nu_{2200})\simeq 3/2\,{\cal A}_2\nu_{2200}\tau_M$ for ${\cal A}_2\nu_{2200}\tau_M
\ll 1$. This indicates that the sin-harmonics may in fact dominate the spectrum of tidal libration. There is, however,
another circumstance that may prevent this. For very low effective viscosities and sufficiently wide orbits, when
$n\,\tau_M\ll 1$, the kvalitet is a slowly declining function of frequency,
\eb
K_s(2,\nu)=\frac{3}{2}\left(1-{\cal A}_2(1+{\cal A}_2)\lambda^2+O(\lambda^4)\right),
\ee
so that $K_s(2,n)\approx K_s(2,0)$. In this case, the pairs of symmetric terms ${q,j}={0,1}$ and {1,0}, and ${0,-1}$ and
${-1,0}$, almost cancel each other, being opposite in sign, and differing by the factor $K_s(2,\nu)$ in magnitude. This
happens to our model Titan with a nominal $\tau_M=0.1$ d considered in the previous simulations. For the sake
of completeness, we should consider another specific scenario when ${\cal A}_2\,n\,\tau_M\gg 1$. Then,
$K_s(2,n)\approx3/2(1+{\cal A}_2)$ and, limiting the expansion in $e$ to the first power,
\eb
\ddot\theta_{\rm sin}\approx 6\frac{M_1}{\xi M_2}n^2\left(\frac{R}{a}\right)^3 \frac{{\cal A}_2}
{1+{\cal A}_2}\,e\,\sin{\cal M}.
\ee
Due to the positive sign of this periodic force, it acts in a similar fashion to the triaxial torque as a restorative
force. A stable pseudosynchronous equilibrium is not possible if the offset of the equilibrium torque in frequency
is smaller that the amplitude of the forced libration caused by this periodic tidal torque, which leads to
\eb
e_{\rm crit}= \frac{M_1}{\xi M_2}\left(\frac{R}{a}\right)^3 \frac{{\cal A}_2}{1+{\cal A}_2}.
\ee
and the only condition of pseudosynchronism is $e>e_{\rm crit}$. For Titan, the calculated value of $e_{\rm crit}$
is a rather low $9\cdot 10^{-5}$. Such small numbers are expected for practically all known star-planet and planet-moon
systems because $R\ll a$. In conclusion, we identified one area of parameters for rotationally symmetric bodies
(${\cal A}_2\,n\,\tau_M\gg 1$) where pseudosynchronous equilibrium can be precluded by the oscillating components of
the tidal torque, but the corresponding threshold eccentricity is so low that such cases should hardly exist.

\section{Summary}
\label{sum.sec}
The rate of energy dissipated in a uniform deformable body by tides in the dominating degree $l=2$ is
\begin{eqnarray}
\langle \frac{dE}{dt}\rangle &=&  \vspace{-5mm}\frac{{\cal G}M_1^2R^5}{a^6}\; \sum_{m=0}^2 \frac{(2-m)!}{(2+m)!}(2-\delta_{0m}) \nonumber \\
&&\times\sum_{p=0}^2 \sum_{q=-\infty}^{+\infty}[F_{2mp}(i)\,G_{2pq}(e)]^2\,\nu_{2mpq}K_c(\nu_{2mpq})
\label{dedt.eq}
\end{eqnarray}
where $i$ is the equator obliquity relative to the orbit plane, $F_{2mp}(i)$ are the inclination functions,
and other parameters are defined elsewhere \citep{heat1}. Due to the presence of the positively definite tidal
frequency $\nu_{2mpq}$, the principal resonant mode of the tidal torque is nullified and does not participate in
the dissipation of kinetic energy. For example, the main component of the torque for small and moderate eccentricities is
the $mpq=200$ term, but it vanishes in Eq. \ref{dedt.eq} if the planet is captured into the 1:1 spin-orbit resonance.
The other tidal modes, separated from the dominant mode
by an integer times $n$, will continue to contribute to dissipation. A semiliquid body caught in pseudosynchronous rotation, on the other hand, continues to
rotate with respect to the perturber's vector in the prograde sense, so that the dominant mode of tidal torque 
continues to generate some dissipation (limited due to the smallness of the tidal frequency $\nu_{2200}$);
however, the negative torque from this mode will be exactly balanced, on average, by the component torques
from the other modes. As a result, the rate of tidal dissipation is {\it smaller} at pseudosynchronism than at
synchronism. The fact that the heating rate is minimal at the point of tidal equilibrium has been known and
discussed in the literature \citep{wis,lev}.

In this paper, only the ``principal" pseudosynchronous state has been considered explicitly, i.e. the one
associated with the 1:1 spin-orbit resonance. This limits the analysis to the case of small and moderate eccentricities.
The same mathematics can be easily generalized for larger eccentricities ($e\ga 0.2$), where equilibrium torque
emerges at other tidal modes. Using the terminology by \citep{sto}, only the true, or resonant, equilibria will be
stable, whereas unstable equilibria of null secular torque can be found between the resonances. These have no practical
significance, and they are not considered in this paper. This may be true only for purely Maxwell-type viscoelastic
models without any secondary inelastic mechanisms of dissipation, however. The action of an Andrade-type dislocation
and unpinning mechanism, for example, coupled with a characteristic low-frequency cutoff \citep{kaspe}, may concievably
lead to locally stable equilibria between the spin-orbit resonances. Observations of the Lunar tidal phase lag with
GRAIL indicate the existence of a secondary peak in the general kvalitet factor $k_2/Q$ between the frequencies
$0.075\,n$ and $1\,n$ \citep{wil1}. This secondary peak can be heuristically fitted with a Debye peak model
\citep{wil2,wil3}, but the Andrade creep model with a $\tau_A\simeq 0.45$ yr can also reproduce this feature.
If the Moon's eccentricity were higher, the secondary peak of the main resonance could be shifted toward positive
values of secular torque by the left shoulder of the 3:2 resonance in such a way that a zero-torque crossing with a
locally negative slope takes place between these two resonances. This should make non-resonant rotational equilibria
possible in principle. A more detailed analysis supported by numerical simulations is required to investigate this curious 
possibility.

We have demonstrated that pseudosynchronous rotation is only possible under specific, restrictive conditions. As long as
the assumption of a homogeneous body is applicable,
it can only happen to rather inviscid, semiliquid or semi-molten bodies with a characteristic tidal wave number
$\lambda=\tau_M n\ll 1$. If this condition is not satisfied, the offset of equilibrium secular torque collapses
and complete synchronization is the only stable state. Secondly, the orbital eccentricity should be large enough
to accommodate the range of free libration to avoid overshooting the resonance at the minimum velocity, per Eq. \ref{ecrit.eq}. The threshold value depends on the degree of elongation $(B-A)/C$, hence, more rotationally symmetric
bodies have better chances to be pseudosynchronous than their more deformed counterparts. On the other hand, eccentricity
may be limited by the first condition for fixed viscosity and shear modulus, see. Eq. \ref{eta.eq}. The combination
of these opposing criteria implies that pseudosynchronous equilibria are practically impossible for large satellites like
Titan if the effective viscosity is greater than $\approx 10^{15}$ Pa$\cdot$s.

One possibly observable aspect of a stable pseudosynchronous state is that the large-amplitude,
long-period free librations of angular velocity around the equilibrium value $\dot\gamma_{\rm pseudo}$ are
eternal, i.e., not diminishing in amplitude with time (cf. Fig. \ref{capture.fig}, right). Physically, this
fact is readily explained by the null work produced by the secular tidal torque in that state. For the free
libration to be damped, the work of secular torque needs to be negative. This happens when the periodic torque
and velocity have opposite phase, whereas in our case, the torque changes its sign over one period of libration
but the velocity $\dot\gamma$ remains positive. On the other hand, the tidal bulge constantly moves across the
surface, the body is being deformed and stressed, and the energy dissipation, being minimized at pseudosynchronism,
still goes on. Obviously, the energy spent on tidal heating is not taken from the kinetic energy of the rotating body,
but directly from the orbital energy. Inevitably, the orbit should gradually shrink, unless it is supported by
a fast rotating perturber. 

Being only marginally stable, a pseudosynchronous body can spontaneously leave this state and fall into the 1:1 resonance
if the eccentricity falls below the critical value.
The tidally driven rate of eccentricity is negative in the vicinity of the 1:1 resonance for small $e$. Most
tidally interacting planets and satellites should therefore be eventually captured in that resonance, unless
the eccentricity is boosted by the tides in the primary or other bodies. This may be the reason we do not find any
pseudosychronously rotating satellites in the Solar system. Titan, with its current value of eccentricity, could
not be pseudosynchronous, but if its orbit was more elongated in the past, it could be pseudosynchronous for
a limited duration. 
\section*{Acknowledgments}
I thank the anonymous referee for the exceptionally thorough and constructive review of the paper. Generous help and support from M. Efroimsky, B. Noyelles, W. Henning, and J. Frouard are also greatly appreciated.

\label{lastpage}

\end{document}